\documentstyle[12pt]{article}
\def\slash#1{\setbox0=\hbox{$#1$}#1\hskip-\wd0\hbox to\wd0{\hss\sl/\/\hss}}

\begin{document}

\begin{flushright}
MZ-TH/92-30 \\
July 1992
\end{flushright}
\bigskip

\begin{center}
{\bf{\LARGE Leptonic {\em CP} Asymmetries in}} \\[0.45cm]
{\bf{\LARGE Flavor-changing $\mbox{\em H}^{\, \mbox{{\large 0}}}$
Decays}} \\[1.5cm]
\bigskip\bigskip\bigskip
{\large J.~G.~K\"orner \footnote[1]{work supported in part by the BMFT,
{\em FRG}, under contract 06MZ730.},
A.~Pilaftsis \footnote[2]{supported by a grant
from the Postdoctoral Graduate College of Mainz.},
K.~Schilcher \footnotemark[1] } \\
 Institut f\"ur Physik \\
 Johannes-Gutenberg-Universit\"at \\
 Staudinger Weg 7, Postfach 3980 \\
 D-6500 Mainz, {\em FRG}
\end{center}

\bigskip
\centerline{ to appear in {\em Physical Review D}}
\bigskip\bigskip
\centerline {\bf ABSTRACT}

Leptonic flavor-changing $H^0$ decays with
branching ratios of the order of $10^{-5}-
10^{-6}$ may constitute an interesting framework when looking for large
$CP$-violating effects. We show that leptonic $CP$~asymmetries
of an intermediate $H^0$ boson can be fairly large in natural scenarios
of the minimal  Standard Model ($SM$) with right-handed neutrinos,
at a level that may be probed at future $H^0$~factories.\\

\newpage

\section*{1.~Introduction}

\indent

Higgs physics has become the subject of many recent theoretical
studies~[1,2,6]. It was recently argued that  the lepton-flavor-violating
decays of an intermediate $H^0$ can be relatively large in the context of
the $SM$ with  one right-handed neutrino field per family~[2]. On the contrary,
ordinary see-saw models predict vanishingly small flavor-changing decay
rates for the $H^0$ or $Z^0$ particle~[3] relying on the strong assumption
that the Dirac and Majorana mass matrices $m_D$ and $m_M$, appearing
in the Yukawa sector, can be simultaneously brought to a diagonal form.
This scenario is equivalent to the single-family case, where the mass
of the light Majorana neutrino $m_{\nu}$ and its mixing angle with the
heavy one $\xi_{\nu N}$ are
\begin{equation}
m_\nu \approx \frac{m_D^2}{m_N} \ , \qquad \quad \xi_{\nu N}\approx
\frac{m_D}{m_M}\approx  \sqrt{\frac{m_\nu}{m_N}}
\end{equation} 
Consequently,  one gets very heavy Majorana neutrinos with
$m_N \stackrel{\displaystyle >}{\sim} 10^7$~GeV  for neutrinos that are
consistent with cosmological constraints and extremely suppressed mixing
angles $\xi_{\nu N} \stackrel{\displaystyle <}{\sim} 10^{-6}$,
if one approximates the Dirac mass matrix $m_D$ by the quark
or charged  lepton mass matrix, as dictated by many $GUT$ models (i.e.~$m_D
\sim
m_{leptons} \sim 1$~GeV). Similar
conclusions are obtained, even if one
right-handed neutrino with Majorana interactions will be added in the three
generation model~[4]. This
situation, however, changes drastically in a two or
three generation model~[5,6].
In general, the Yukawa sector containing both Dirac and Majorana terms can
be represented in an appropriate Majorana basis of the neutrino
fields as follows:
\begin{equation}
-{\cal L}_M^\nu \ = \ \frac{1}{2} (\bar{\nu}^0_L, \bar{\nu}_R^{0C})
\left( \begin{array}{cc}
0     & m_D\\
m_D^T & m_M \end{array} \right)
\left( \begin{array}{c}
\nu_L^{0C} \\ \nu_R^0 \end{array} \right)\quad + \quad h.c.
\end{equation} 
where one can always assume  $m_M$ to be a real diagonal $n_R \times n_R$
matrix and $m_D$ an arbitrary non-hermitian $n_L \times n_R$
matrix. The parameter $n_L(n_R)$ indicates the number of left(right)-handed
neutrino fields.
Actually, it was demonstrated in~[5] that e.g.~for $n_L=n_R=2$, matrices
of the form
\begin{equation}
m_D \simeq \left( \begin{array}{cc}
a & b \\
c & \displaystyle\frac{bc}{a} \end{array} \right)\ , \qquad \quad
m_M \simeq A \left( \begin{array}{cc}
1  &  0 \\
0 & \displaystyle -\frac{b^2}{a^2} \end{array} \right)
\end{equation} 
lead automatically to two approximately massless neutrinos.
If, for instance, $a=b=c$, we then obtain the two generation
version of a democratic family mixing model.
The two acossiated heavy neutrinos have masses \\
\begin{equation}
m_{N_1} \approx A \qquad\qquad m_{N_2} \approx \frac{b^2}{a^2} A
\end{equation} 
The only constraints on the free parameters $a,\ b,\ c,\ A$ are imposed
by phenomenology. In a global analysis~[7] based on charge-current
universality,
neutral current effects etc. it was found, for example, that
the allowed maximum value for ($\xi\xi^\dagger)_{ij}$ is of the order
of 0.01--0.13, where the larger value refers to the heavy--light mixings
of the systems $e-\tau$ or $\mu - \tau$ and the lower value to mixing in
the $e-\mu$~sector.\\

In this work we wish to investigate whether $CP$ asymmetries
defined as
\begin{equation}
A_{CP}^{ij} \ = \ \frac{\Gamma (H^0 \to l_i\bar{l}_j)\ -\
\Gamma (H^0 \to \bar{l}_il_j)}{\Gamma (H^0 \to l_i\bar{l}_j)\ +\
\Gamma (H^0 \to \bar{l}_il_j)}
\end{equation} 
can, {\em in principle}, be observed at $LHC$ or $SSC$ energies,
with $l_i$ being charged leptons. We will show that such $CP$ effects
are indeed sizeable for the $e-\tau$ or $\mu -\tau$~system, if the minimal $SM$
is extended by three right-handed neutrino fields,
and would be experimantally measurable when a high
efficiency in the $\tau$-lepton identification is achieved.\\

\section*{2.~The \boldmath $SM$ \unboldmath with right-handed neutrinos}

\indent

Although the case of an equal number of left-handed and
right-handed neutrino states may seem more aesthetical, in general the number
of right-handed neutrinos that can be added in the
$SM$ is arbitrary~[8].
Thus, the symmetric neutrino mass matrix $M^\nu$ given in eq.~(2) can
generally be diagonalized by a $(n_L+n_R)\times (n_L+n_R)$ unitary
matrix $U^\nu$ in the following way:
\begin{equation}
U^{\nu T} M^\nu U^\nu \ = \ \hat{M}^\nu
\end{equation} 
This gives $n_L$ light neutrino mass eigenstates ($\nu_i$) and $n_R$
heavy ones $(N_i)$ which are related to the weak eigenstates via
\begin{equation}
\left( \begin{array}{c}
\nu^{0C}_L \\ \nu_R^0 \end{array} \right) \ = \
U^\nu \left( \begin{array}{c}
\nu_R \\ N_R \end{array} \right)\ , \qquad \quad
\left( \begin{array}{c}
\nu^{0}_L \\ \nu_R^{0C} \end{array} \right) \ = \
U^{\nu\ast} \left( \begin{array}{c}
\nu_L \\ N_L \end{array} \right)\
\end{equation} 
Denoting the neutrino mass eigenstates by $n_i$ (i.e.~$n_i\equiv \nu_i$
for $i=1,2,\dots , n_L$ and $n_i\equiv N_{i-n_L}$ for $i=n_L+1,\dots ,
n_L+n_R$), we can write down the relevant Lagrangians that describes the
interactions between Majorana neutrinos $n_i$ and $W^\pm$ or $H^0$
bosons~[6]. One has
\begin{eqnarray}
{\cal L}_{int}^W& = & -\frac{g_W}{2\sqrt{2}} W^{-\mu}
{\bar{l}}_i \ B_{l_ij} {\gamma}_{\mu} (1-{\gamma}_5) \ n_j \quad + \quad h.c.
\\[0.3cm]
{\cal L}_{int}^H & = & - \frac{g_W}{4M_W} H^0
{\bar{n}}_i \ [(m_{n_i}+m_{n_j})\mbox{Re}(C_{ij})+
i{\gamma}_5(m_{n_j}-m_{n_i})\mbox{Im}(C_{ij})] \ n_j
\end{eqnarray} 
where $B$ and $C$ are $n_L\times (n_L+n_R)$ and $(n_L+n_R)\times (n_L+n_R)$
dimensional matrices, respectively, which are defined as
\begin{eqnarray}
B_{l_ij} & = & \sum\limits_{k=1}^{n_L} V^l_{l_ik} U^{\nu\ast}_{kj}
\quad  \quad \mbox{with} \quad j=1,\dots,n_L+n_R \\
C_{ij} & = & \sum\limits_{k=1}^{n_L} U^{\nu}_{ki} U^{\nu\ast}_{kj}
\quad \quad \mbox{with} \quad i,j=1,2,\dots,n_L+n_R
\end{eqnarray} 
$V^l$ in eq.~(10) is the relevant $n_L\times n_L$ Cabbibo-Kobayashi-Maskawa
($CKM$) matrix. For completeness, we also give the charged current
Lagrangian, which describes the coupling of the Majorana neutrinos $n_i$
with the unphysical Goldstone bosons $\chi^\pm$ in the Feynman--'t Hooft
gauge~[6].
\begin{eqnarray}
{\cal L}_{int}^{\chi}\  &=&\   -\frac{g_W}{2\sqrt{2}M_W} {\chi}^-
{\bar{l}}_i \ [m_{l_i}B_{l_ij} (1-{\gamma}_5)\ -\ B_{l_ij}(1+{\gamma}_5)
m_{n_j}] \ n_j\quad + \quad h.c. \nonumber\\
\end{eqnarray} 

Some coments on the structure of $U^\nu$ are in order. A sufficient
condition for the $n_L$ light neutrinos to be approximately massless at the
tree level is
\begin{equation}
m_D\ m_M^{-1}\ m_D^T\ \  = \ \ {\bf 0}
\end{equation} 
As already mentioned in the introduction,
eq.~(13) cannot be satisfied by ordinary see-saw
models for finite Majorana mass terms (i.e.~$n_R=1$). This restriction
can naturally be realized by more than one generation. Especially, one
can prove that once condition (13) is valid, $M^\nu$ can be diagonalized
by a unitary matrix $U^\nu$ of the form
\begin{equation}
U^\nu = \ \left( \begin{array}{cc}
(1+\xi^\ast \xi^T )^{-\frac{1}{2}} & \xi^\ast(1+\xi^T\xi^\ast
)^{-\frac{1}{2}}\\
-\xi^T(1+\xi^\ast \xi^T )^{-\frac{1}{2}}&(1+\xi^T \xi^\ast )^{-\frac{1}{2}}
\end{array} \right)
\left( \begin{array}{cc}
{\bf 1} & 0\\
0 & V^N \end{array} \right)
\end{equation} 
where $\xi=m_Dm_M^{-1}$ and $V^N$ is a unitary $n_R\times n_R$ matrix
that diagonalizes the following symmetric matrix:
\begin{equation}
M_M\ = \ \frac{1}{2} (1+\xi^\dagger \xi )^{-\frac{1}{2}} m_M
(1+\xi^T \xi^\ast)^{\frac{1}{2}}
+ \frac{1}{2} (1+\xi^\dagger \xi )^{\frac{1}{2}} m_M
(1+\xi^T \xi^\ast )^{-\frac{1}{2}}
\end{equation} 
Note that the square root of a matrix $A$ is defined as the matrix
$B$ obeying the equality: $BB=A$, where all elements $B_{ij}$ are taken to
be on the first sheet. It should be also noted that the
representation~(14) in terms of the matrix-valued parameter $\xi$ is only
possible for $\|\xi\| <1$. This inequality is obviously consistent with
phenomenological constraints.\\

The matrices $B$ and $C$ satisfy a
number of useful identities. Below we list the most relevant for us, i.e.
\begin{eqnarray}
\sum\limits_{k=1}^{n_L} B_{l_kj}B_{l_ki}^{\ast} & = & \ C_{ij} \\
\sum\limits_{i=1}^{n_L+n_R} m_{n_i}B_{l_1i}B_{l_2i} & = & \ 0 \\
\sum\limits_{i=1}^{n_L+n_R} m_{n_i} B_{li}C^{\ast}_{ij} & = & \ 0 \\
\sum\limits_{i=1}^{n_L+n_R} B_{li}C_{ij} & = & \ B_{lj} \\
\sum\limits_{i=1}^{n_L+n_R} B_{l_1i}B_{l_2i}^{\ast} & = & \ {\delta}_{l_1l_2}
\end{eqnarray} 
Eq.~(20), for example, is the generalized form of the unitarity
condition for $n_L$ charged leptons and $n_L+n_R$ neutrinos.
Since the explicit form of $B$ and $C$ is rather complicated, we can, as
usual, approximate $B$ and $C$ in terms of the mixing parameters $\xi_{\nu N}$.
Up to leading order in $\xi$, these matrices are given by
\begin{eqnarray}
B_{l\nu} = V^l_{l\nu} \ , \quad B_{lN} = (V^l\xi V^{N\ast})_{lN} \\
C_{\nu \nu} = 1 \ , \quad C_{\nu N} = (\xi V^{N\ast})_{\nu N} \ , \quad
C_{NN} = (V^N {\xi}^{\dagger} \xi V^{N\ast})_{NN}
\end{eqnarray} 

Let us now count the number of $CP$ phases in $SU(2)_L\otimes U(1)_Y$
theories with arbitrary number of right-handed neutrinos.
It is convienient to discuss this problem in a weak basis where
$m_M$ is real and diagonal and the charged lepton fields
are rotated so as to coincide with their masseigenstates, i.e.
\begin{equation}
l^0_{L_i} \ = \ V^{l\dagger}_{ij} l_{L_j}
\end{equation} 
At the same time, the neutrino fields have to be transformed as
\begin{equation}
 {\nu^0_{L_i}}' \ =\ V^l_{ij} \nu^0_{L_j}
\end{equation} 
so that the charged current sector is diagonal in this basis. The
number of $CP$~phases equals the number of non-trivial indepedent phases
existing in the non-hermitian $n_L\times n_R$ Dirac mass matrix $m_D$.
Thus, $m_D$ possesses $n_Ln_R$ phases from which only $n_L$ phases coming
from ${\nu^0_{L_i}}'$ can be redefined away, since the phases of $\nu^0_{R_i}$
are fixed by the above assumptions. As a result, the net number of $CP$
phases in these theories will be given by the relation
\begin{equation}
N_{CP}\ =\ n_L(n_R-1)\ , \qquad \ \ \mbox{for} \ \ n_R \geq 1
\end{equation} 
The same conclusion can be derived by following line of arguments similar
to that of ref.~[9]. From eq.~(25) it is obvious that models
with one right-handed neutrino and arbitrary number of left-handed ones~[4]
cannot account for possible $CP$-violating phenomena in the leptonic sector.
In other words, at least two or more right-handed neutrinos  are required
to explain possible $CP$ asymmetries in the leptonic $H^0$ decays.\\

In the next section we will present the analytical calculation and
numerical results for the off-diagonal leptonic $H^0$ decays and
their acossiated $CP$ phenomena. We will discuss these effects in two
illustrative scenarios, where two and three right-handed neutrinos
are present, respectively.\\

\section*{3.~Leptonic \boldmath $CP$ \unboldmath asymmetries in
\boldmath $H^0$ \unboldmath decays}

\indent

As is well known, $CP$ violation requires the existence of at
least two amplitudes with different absorptive parts as well as
different relative  weak phases that cannot be rotated away.
In this context, numerous
studies on possible $CP$ mechanisms that may take place at high energies
and can give rise to large $CP$-violating phenomena have been presented in
the literature over the last years~[10-13]. The above requirement
leads to the additional restriction that at least one heavy Majorana
neutrino $N_1$ with mass
\begin{equation}
m_{N_1}\ \simeq \ 100\ \mbox{GeV}
\end{equation} 
must be present. Then, the effective $l_1-l_2-H^0$ coupling
(see fig.~1) -- with $l_1,\ l_2$ being two different charged letpons --
acquire absorptive contributions for $m_{N_1} < M_H $, which result from
on-shell $\nu_i N_1$~intermediate states as shown in the diagrams~1a and~1b.
Here, of course, we have implicitly assumed that we are dealing with an
intermediate mass Higgs boson (i.e.~$100< M_H \leq 140$~GeV), whose
branching ratio into $b\bar{b}$ pairs is of the order one.
In addition, in order to avoid excessive complication in our calculations, we
make the following realistic asumptions for the vertex function:
\begin{equation}
 m_{l_1}\ =\ 0 \ , \qquad \qquad \frac{m^2_{l_2}}{M^2_W}\  \ll \ 1
\end{equation} 
In fact, the matrix element of the amplitude $H^0 \to \bar{l}_1l_2$
can be parametrized as follows:
\begin{equation}
\mbox{T}(H^0\to \bar{l}_1l_2)\ =\ \frac{g_W\alpha_W}{16\pi} \
\frac{m_{l_2}}{M_W}
\left[ F^{dis}(M^2_H) + iF^{abs}(M^2_H) \right]
\ \bar{\mbox{u}}_{l_2} (1-\gamma_5)\mbox{v}_{l_1}
\end{equation} 
where $F^{dis}(q^2)$ and $F^{abs}(q^2)$ are complex form factors, which
are related
to each other by a subtracted dispersion relation of the form
\begin{equation}
F^{dis}(M^2_H)\ \ =\ \ F^{dis}(q^2=0)\ +\ \frac{M^2_H}{\pi}
\, \, {\cal P}\int\limits_{0}^{\infty}dq^2 \frac{F^{abs}(q^2)}{q^2(q^2-M^2_H)}
\end{equation} 
In what follows, we will focus our attention on two class of models, which
have different phenomenological features.\\

\subsection*{3.1 The model with \boldmath $n_R=2$\unboldmath}

\indent

In this first scenario we assume the presence of two right-handed neutrinos
in the Yukawa sector, i.e. $n_R=2$. In particular, to satisfy condition~(26),
we assume that the resulting two heavy neutrinos
$N_1$ and $N_2$ have masses $m_{N_1}\simeq 100$~GeV and $m_{N_2} >
2M_W$. Since in this model $\mbox{det}\, M^\nu = 0$, one of the
light neutrinos will be massless (at the tree level),
while the other two neutrinos can be
taken to be extremely light by taking condition~(13) to be
approximately valid.\\

The number of mixing parameters $B_{lj}$ and $C_{ij}$ in these theories
seems to be rather large. However, employing identities~(17) and~(18)
we can derive some helpful relationships among them such as
\begin{eqnarray}
|B_{l_2N_2}B_{l_1N_2}^\ast|\ &=&\ \frac{m_{N_1}}{m_{N_2}}
|B_{l_2N_1}B_{l_1N_1}^\ast|\\
B_{l_2N_2}C_{N_2N_2}B_{l_1N_2}^\ast\ &=&\ \frac{m^2_{N_1}}{m^2_{N_2}}
B_{l_2N_1}C_{N_1N_1}B_{l_1N_1}^\ast  \\
|B_{l_2N_1}C_{N_1N_2}B_{l_1N_2}^\ast|\ &=&\
|B_{l_2N_2}C_{N_2N_1}B_{l_1N_1}^\ast|\ =\ \frac{m_{N_1}}{m_{N_2}}
|B_{l_2N_1}C_{N_1N_1}B_{l_1N_1}^\ast|
\end{eqnarray} 
from which we can estimate the order of magnitude for the different
$CKM$-mixing combinations. Actually, we find that expressions
of the form~(31) and~(32) arising from the graphs~(1a) and~(1b) are
$\xi^4$-suppressed and will thus not be considered here. Eq.~(30)
shows a direct relation between heavy neutrino masses and $CKM$
mixings which must also be taken into account. As a
consequence, the form factor $F^{dis}$ evaluated at $q^2=0$~[2] can be cast
into the following form:
\begin{eqnarray}
F^{dis}(q^2=0)\ &=&\ \frac{1}{2} \sum\limits_{i=1}^{n_R=2}
B_{l_2N_i}B_{l_1N_i}^\ast\lambda_i \Bigg\{
-\frac{3}{2}\ \frac{1}{1-\lambda_i}\ +\ \frac{3}{(1-\lambda_i)^2}\ \nonumber\\
&&+\ \frac{3\lambda_i\ln \lambda_i}{(1-\lambda_i)^3}\ +\
\frac{M^2_H}{M^2_W} \bigg[ \frac{\lambda_i\ln \lambda_i}{(1-\lambda_i)^3}\
-\ \frac{1}{2}\ \frac{\lambda_i^2\ln \lambda_i}{(1-\lambda_i)^3}\nonumber\\
&&+\ \frac{3}{4}\ \frac{1}{(1-\lambda_i)^2}\ -\ \frac{1}{4}\ \frac{\lambda_i}
{(1-\lambda_i)^2} \bigg] \Bigg\}
\end{eqnarray} 
where
\begin{equation}
\lambda_i\ =\ \frac{m^2_{N_i}}{M^2_W}\ , \qquad \mbox{for} \qquad i=1,2
\end{equation} 
To make use of eq.~(29), we need the analytical form of $F^{abs}(q^2)$,
which has been calculated by applying the usual
Cutkosky rules~[14]. Keeping terms of order $\xi^2$, we get
\begin{eqnarray}
F^{abs}(q^2)\ &=&\ \sum\limits_{i=1}^{n_R=2}
B_{l_2N_i}B_{l_1N_i}^\ast\lambda_i \Bigg\{
-I(q^2,m^2_{N_i},0)\ +\ K(q^2,m^2_{N_i},0,M^2_W)\nonumber\\
&&-\ 4F(q^2,m^2_{N_i},0,M^2_W)\ +\ 2F(q^2,M^2_W,M^2_W,m^2_{N_i})\nonumber\\
&&+\ 2\left( 2+\frac{q^2}{M^2_W} \right) G(q^2,M^2_W,M^2_W,m^2_{N_i})\
-\ 2K(q^2,M^2_W,M^2_W,m^2_{N_i})\nonumber\\
&&+\ \frac{M^2_H}{M^2_W}\left[ F(q^2,M^2_W,M^2_W,m^2_{N_i})\ -\
K(q^2,M^2_W,M^2_W,m^2_{N_i}) \right] \Bigg\}
\end{eqnarray} 
where the functions $I,\ F,\ K,\ G$ are given below
\begin{eqnarray}
I(q^2,m^2_1,m^2_2)\ &=&\ \theta(q^2-(m_1+m_2)^2)\frac{\pi}{2q^2}
\lambda^{1/2}(q^2,m^2_1,m^2_2)\nonumber\\
\mbox{with} \qquad \quad\quad  \lambda (x,y,z)\ &=&\ (x-y-z)^2-4yz\\[0.5cm]
K(q^2,m^2_1,m^2_2,M^2)\ &=&\ \theta(q^2-(m_1+m_2)^2) \frac{\pi}{2}\
\frac{M^2_W}{q^2}\ \ln\left[ \frac{t^+(q^2,m^2_1,m^2_2)-M^2}{
t^-(q^2,m^2_1,m^2_2)-M^2} \right]\nonumber\\
\mbox{with}\qquad \quad t^\pm\ &=&\ -\frac{1}{2}\left[ q^2-m^2_1-m^2_2
\mp\lambda^{1/2}(q^2,m^2_1,m^2_2) \right]\\[0.5cm]
F(q^2,m^2_1,m^2_2,M^2)\ &=&\ \frac{M^2_W}{q^2}I(q^2,m^2_1,m^2_2)\ +\
\frac{M^2-m^2_1}{q^2}K(q^2,m^2_1,m^2_2,M^2)\\[0.5cm]
G(q^2,m^2_1,m^2_2,M^2)\ &=&\ \frac{M^2_W}{M^2} \left[
F(q^2,m^2_1,m^2_2,M^2)\ -\ F(q^2,m^2_1,m^2_2,0) \right]
\end{eqnarray} 
Finally, the absorptive part of $F(M^2_H)$ for $H^0$-mass values being
above the $N_1$ mass and below the threshold of real $WW$ production is
obtained by
\begin{eqnarray}
F^{abs}(M^2_H)\ &=&\ B_{l_2N_1}B_{l_1N_1}^\ast\frac{m^2_{N_1}}{M^2_W}
\bigg[ -I(M^2_H,m^2_{N_1},0)\ +\ K(M^2_H,m^2_{N_1},0,M^2_W)\nonumber\\
&&-\ 4F(M^2_H,m^2_{N_1},0,M^2_W) \bigg]
\end{eqnarray} 
Being optimistic in our numerical considerations, we use the value
0.13 for the $CKM$-mixing combination $|B_{l_2N_1}B_{l_1N_1}^\ast|$
which is about the maximal experimentally
allowed value for $l_2=\tau$ and $l_1=e$ or $\mu$~[15].
So, after performing numerically the integration in eq.~(29), we
evaluate the partial width of the $H^0$ decay into two different leptons
through the expression
\begin{eqnarray}
\Gamma(H^0\to \bar{l}_1l_2\ \ \mbox{or}\ \ l_1\bar{l}_2)\ &=& \
\frac{\alpha_W^3}{128\pi^2}\ \frac{m^2_{l_2}}{M^2_W}\ M_H
\bigg[ \left| F^{dis}(M^2_H) \right|^2
\ +\ \left| F^{abs}(M^2_H) \right|^2 \bigg]\nonumber\\
&&
\end{eqnarray} 
Eq.~(41) also enable us to check quantitatively the correctness of our formulas
for sufficient small values of the ratio $M^2_H/4M^2_W$. We have found them to
be in agreement with the numerical results given in~[2].\\

We are now in the position to investigate numerically the observability of
the off-diagonal leptonic decay modes of $H^0$ by studying the ratio
\begin{equation}
R_{l_1l_2}\ =\ \frac{\Gamma(H^0\to \bar{l}_1 l_2)\ +\
\Gamma(H^0\to l_1\bar{l}_2)}{\Gamma(H^0\to b\bar{b})}
\end{equation} 
The dominant decay channel of $H^0$ to $b\bar{b}$ pairs, after
including $QCD$ corrections, has been estimated in~[16]. The results of~[16]
can be summarized as follows:
\begin{eqnarray}
\Gamma (H^0 \to b\bar{b} ) = \frac{3{\alpha}_W}{8} M_H
\frac{{\bar{m}}_b^2}{M^2_W} (1 + 1.8{\alpha}_s +2.95{\alpha}_s^2)
\left( 1+ {\cal O}(\frac{{\bar{m}}_b^2}{M^2_W}) \right)
\end{eqnarray} 
where $\bar{m}_b$ is the running b-quark mass given by
\begin{equation}
{\bar{m}}_b = {\hat{m}}_b \left( \frac{23{\alpha}_s}{6\pi} \right)^{
\frac{12}{23}} (1+0.374{\alpha}_s+0.15{\alpha}_s^2) \,
\end{equation}
with $\hat{m}_b= 8.23$~GeV, for $\Lambda_{\overline{MS}}=150$~MeV.
The $QCD$ coupling $\alpha_s$ and running mass $\bar{m}_b$ are
evaluated at scale $M_H$.\\

In this model the $CP$-asymmetry parameter $A_{CP}$
defined in eq.~(5) turns out to be
\begin{equation}
A_{CP} \ \simeq\ 2\sin \delta_{CP}\
\frac{F^{abs}(M^2_H)F^{dis}_{N_2}(M^2_H)}
{ | F^{abs}(M^2_H)|^2\ +\ | F^{dis}_{N_1}(M^2_H)|^2}
\end{equation} 
where $\delta_{CP}$ is a $CP$-odd rephasing-invariant angle
between the $CKM$ expressions  $B_{l_2N_1}B_{l_1N_1}^\ast$ and
$B_{l_2N_2}B_{l_1N_2}^\ast$.
Unfortunately, eq.~(31) implies that $\delta_{CP}=0$ and hence $A_{CP}=0$.
In eq.~(45) $F^{dis}_{N_i}(M^2_H)$ stands for the
$i$th summation term in eq.~(33).
Because of relation~(30) $F^{dis}_{N_1}(M^2_H)$ will give the biggest
contribution to $F^{dis}$. Therefore the branching ratio
will be rather indepedent of the heavy neutrino mass~$N_2$.
In table 1 we present the numerical results
for the branching ratio $R$ as function of the $N_2$ mass.
On the other hand, as has been discussed extensively in~[17], one expects
that $LHC$ or $SSC$
colliders will produce $10^6-10^7$ $H^0$ particles
a year provided $M_H \leq 140$~GeV. Therefore, in the $SM$
with two right-handed neutrinos, flavor-changing decays of $H^0$
in the leptonic sector
could, {\em in principle}, be observed if a next generation luminosity
upgrade of a factor of 10  is assumed.\\

\subsection*{3.2 The model with  \boldmath $n_R=3$ \unboldmath}

\indent

Let us now consider the case where each left-handed neutrino $\nu^0_{L_i}$
has a right-handed partner $\nu^0_{R_i}$. This scenario is more
symmetric than the previous one and may hence represent a rather
realistic situation.
In particular, we assume that the one heavy Majorana neutrino $N_1$ is
relatively light, i.e.~$m_{N_1} \simeq 100$~GeV, while the other two
neutrinos are non-degenerate (i.e.~$m_{N_1} \ll m_{N_2} < m_{N_3}$), but their
masses lie in the TeV range.\\
{}From eq.~(18) we have the folowing useful approximations:
\begin{eqnarray}
C_{N_3N_3}\ &\simeq &\ \frac{m^2_{N_2}}{m^2_{N_3}}\
\frac{|B_{l_2N_2}|^2}{|B_{l_2N_3}|^2}\ C_{N_2N_2}\\
C_{N_3N_2}\ &\simeq &\ -\ \frac{m_{N_2}}{m_{N_3}}\
\frac{B_{l_2N_2}^\ast}{B_{l_2N_3}^\ast}\ C_{N_2N_2}
\end{eqnarray} 
As a result of eqs (46) and (47), the contribution to the form factor
$F^{dis}(q^2)$ coming from the mixing angle quantity
$|B_{l_2N_2}C_{N_2N_2}B_{l_1N_2}^\ast|$ may become the most dominant one.
To leading order in an expansion of
$1/\lambda_2$ we get~[2]
\begin{equation}
F^{dis}(q^2=0)\ \simeq\ \frac{3}{4}B_{l_2N_2}C_{N_2N_2}B_{l_1N_2}^\ast\
\lambda_2
\end{equation} 
For the  phenomenologically compatible values of $C_{N_2N_2}
\stackrel{\displaystyle <}{\sim} 0.1$ we readily see that the other
$CKM$-mixing terms appearing in eq.~(48) can safely be neglected provided
$m_{N_{2,3}} \stackrel{\displaystyle >}{\sim} 1$~TeV. In this scenario the
ratio $R_{l_1l_2}$ behaves like
\begin{equation}
R_{l_1l_2} \simeq 12 \left( \frac{{\alpha}_W}{32\pi} \right)^2
\frac{m_{l_2}^2}{{\bar{m}}_b^2}\ \lambda_2^2\
|B_{l_2N_2}C_{N_2N_2}B_{l_1N_2}^\ast|^2
\end{equation} 
while the $CP$ asymmetry can be simply obtained by
\begin{equation}
A_{CP} \ \simeq\ \frac{8}{3} \ \frac{\sin \delta_{CP}}
{|B_{l_2N_2}C_{N_2N_2}B_{l_1N_2}^\ast|}\
\frac{F^{abs}(M^2_H)}{\lambda_2}
\end{equation} 

Before proceeding with the presentation of the numerical results, we
wish to comment on the dramatic enhancement of the quantity $R$ by
increasing the heavy neutrino mass in eq.~(49). For definiteness, when
$m_{l_2}=m_\tau$, we find that
\begin{equation}
R^{max}_{e\tau}, R^{max}_{\mu \tau} \ \simeq \ \ 10^{-2}
\end{equation} 
for $m_{N_2}=10$~TeV and $B_{l_2N_2}C_{N_2N_2}B_{l_1N_2}^\ast \simeq 0.01$.
Here two different constraints must be taken into account:
the requirement of pertubative unitarity and consistency with
the existing data on lepton-flavor-changing $Z^0$ decays. The former
restriction gives roughly an upper bound
\begin{equation}
\xi^2_{\nu N} \ \leq\ \frac{8}{\alpha_W}\ \frac{M^2_W}{m^2_N}\
\simeq\ 0.03
\end{equation} 
for $m_{N_2} \simeq 10$~TeV
by imposing the inequality condition $\Gamma_N/m_N \leq 1/2$, where
$\Gamma_N$ is the width of the heavy neutrino $N$. Nevertheless, to
the best of our knowledge, constraints from leptonic $Z^0$ decays have not been
considered as yet in this class of models. Adapting the results for
$Z^0 \to b\bar{s}$ of ref.~[18], we get a first estimate for
the $Z^0 \to l_1\bar{l}_2$ matrix element:
\begin{eqnarray}
\mbox{T}(Z^0\to l_1\bar{l}_2)\ \sim\ \frac{g_W\alpha_W}{32\pi\cos \theta_W}
B_{l_2N_2}C_{N_2N_2}B_{l_1N_2}^\ast \frac{m^2_{N_2}}{M^2_W}\
\bar{\mbox{u}}_{l_2} \gamma_\mu(1-\gamma_5)\mbox{v}_{l_1}\ \varepsilon^\mu_Z
\nonumber\\
\end{eqnarray} 
In order to reconcile the experimental value of~[19]
\begin{equation}
\mbox{BR}(Z^0\to e\tau\ \mbox{or}\  \mu\tau) \ \leq\ 10^{-5}
\end{equation} 
with the theoretical prediction, we must constrain mixing-angles and heavy
neutrino masses to the following range:
\begin{equation}
|B_{l_2N_2}C_{N_2N_2}B_{l_1N_2}^\ast|^2\frac{m^4_{N_2}}{M^4_W}\
\leq\  10^3
\end{equation} 
Consequently, in order to be consistent with eqs~(52) and~(55) in our
numerical estimates, we use the optimistic value of $\sim 0.001$
for $B_{l_2N_2}C_{N_2N_2}B_{l_1N_2}^\ast$. From table 2 we see that
this scenario yields rather encouraging branching-ratio values of the
order of $10^{-4}-10^{-5}$ and $CP$ asymmetries of the order of one
($\sim 40\ \%$).\\

\section*{4.~Conclusions}

\indent

Both flavor-violating signals and $CP$ violation in leptonic $H^0$ decays may
arise in the Standard Model with three right-handed neutrinos,
at a level that may be tested at future $H^0$~factories. Such
class of models can naturally provide, tiny masses for the light neutrinos,
relatively light masses
(e.g.~of the order of 100 GeV) for the heavy ones, as well as large
light-heavy neutrino mixings (i.e. $\xi_{\nu N} \sim 0.1$),
contrary to the traditional see-saw models. We find that these
models may reveal attractive phenomenological features at high energy
colliders such as $SSC$ or $LHC$. As has been shown, branching ratios of
the order of $10^{-4}-10^{-5}$ and $CP$ asymmetries of order one in
leptonic off-diagonal $H^0$ decays can, {\em in principle}, be detected in
these planned collider machines. More precisely, we suspect to be able to
analyze $10^2-10^3$~events for the decays $H^0 \to e\tau$ or $\mu\tau$.
Then, the contributing background to the above decays will be about
$10^4$ events -- in which the Higgs invariant mass has been reconstructed
by using leptonic subsequent decays of $t\bar{t}$ pairs~[20] -- which is
10--$10^2$ times bigger than the desired signal. However, the simultaneous
knowledge of large CP~asymmetries of the order of~$10^{-1}-10^{-2}$
(after including the background mentioned above) makes it theoretically
possible to establish such small and very interesting effects in the
leptonic $H^0$~decays.
Finally, special attention has been paid to the fact that
the resulting values of the phenomena discussed in this work are
consistent with all the existing information arising
from neutrino-mixing effects, the validity of pertubative
unitarity and data of rare $Z^0$ decays at $LEP$.\\

In conclusion, in the simplest model which predicts heavy Majorana
neutrinos, i.e.~the Standard Model with right-handed neutrinos, we have
explicitly shown that {\em flavor nonconservation and $CP$ violation in
leptonic $H^0$ decays can indeed be sizeable and are generally not
suppressed by the usual see-saw mechanism}. The attractive theoretical
aspects of this minimal class of models may also lead to further
phenomenological investigations in the gauge sector of the $SM$ or/and
in the kaonic system~[21].\\[1.cm]
{\bf Note added.} The decay processes $Z^0\to l_1\bar{l}_2$
have recently been analyzed in~[22] in the context of the models
discussed here, giving branching ratios closed
to those that are qualitatively obtained by eqs~(53) and~(55).

\newpage

\centerline{\bf\Large Figure and Table Captions }
\vspace{1cm}
\newcounter{fig}
\begin{list}{\bf\rm Fig. \arabic{fig}:}{\usecounter{fig}
\labelwidth1.6cm \leftmargin2.5cm \labelsep0.4cm \itemsep0ex plus0.2ex}

\item Feynman graphs responsible for the flavour-changing lepton--lepton--$H^0$
vertex in the Feynman--'t Hooft gauge. The dashed lines indicate on-shell
contributions from the intermediate states.
\end{list}
\newcounter{tab}
\begin{list}{\bf\rm Tab. \arabic{tab}:}{\usecounter{tab}
\labelwidth1.6cm \leftmargin2.5cm \labelsep0.4cm \itemsep0ex plus0.2ex}

\item  Numerical results of the ratio $R^{\max}$
for the decay modes $H^0 \to e \tau$ or $\mu \tau$ in the
SM with two right-handed neutrino fields ($n_R=2$).
In our numerical estimates we have used the following set of parameters:
$M_W=80.6$~{\rm GeV}, $M_Z=91.161$~{\rm GeV},
$m_{\tau}=1.784$~{\rm GeV}, ${\alpha}_{em}(M^2_W)=1/128$,
$B_{l_2N_1}B_{l_1N_1}^\ast \simeq 0.13$.

\item Numerical results for $R^{max}$ and $A_{CP}$ in a model whith
three right-handed neutrinos. Here, we have
$m_{N_2} < m_{N_3}$ and $B_{l_2N_2}C_{N_2N_2}B_{l_1N_2}^\ast \sim 0.001$.

\end{list}

\newpage
\bigskip\bigskip\bigskip\bigskip\bigskip\bigskip
\centerline{\bf\Large Table 1}
\vspace{1.5cm}
\begin{tabular*}{10.49cm}{|r|c|c|c}
\hline
 & & & \\
$m_{N_2}$ & \hspace{1.5cm} $R^{max}_{l\tau}$\hspace{1.5cm}  & \hspace{1.5cm}
$R^{max}_{l\tau}\qquad \qquad$ &\\
$[$TeV$]$ & $n_R=2$ & $n_R=2$ & \\
  & $M_H=120$~GeV & $M_H=140$~GeV &  \\
& & &\\
\hline\hline
&&& \\
0.2 & 1.42~$10^{-7}$ & 2.27~$10^{-7}$& \\
0.5 & 1.39~$10^{-7}$ & 2.26~$10^{-7}$& \\
0.7 & 1.38~$10^{-7}$ & 2.33~$10^{-7}$& \\
1   & 1.40~$10^{-7}$ & 2.31~$10^{-7}$& \\
1.5 & 1.39~$10^{-7}$ & 2.27~$10^{-7}$& \\
5   & 1.37~$10^{-7}$ & 2.23~$10^{-7}$& \\
&&& \\
\hline
\end{tabular*}

\newpage
\bigskip\bigskip\bigskip\bigskip\bigskip\bigskip
\centerline{\bf\Large Table 2}
\vspace{1.5cm}
\begin{tabular*}{14.34cm}{|r|rr|rr|}
\hline
 & & & & \\
$m_{N_2}$ &  $R^{max}_{l\tau}$  & $R^{max}_{l\tau}$ & $|A_{CP}|$ & $|A_{CP}|$
\\
$[$TeV$]$ & $n_R=3$ & $n_R=3$ & & \\
  & $M_H=120$~GeV & $M_H=140$~GeV &  $M_H=120$~GeV & $M_H=140$~GeV \\
& & & &\\
\hline\hline
&&&& \\
1 & 1.6~$10^{-8}$ & 1.8~$10^{-8}$ & 0.80 & 0.90 \\
2 & 2.2~$10^{-7}$ & 2.4~$10^{-7}$ & 0.70 & 0.80 \\
3 & 1.1~$10^{-6}$ & 1.2~$10^{-6}$ & 0.50 & 0.70 \\
5 & 8.2~$10^{-6}$ & 8.5~$10^{-6}$ & 0.22 & 0.27 \\
7 & 3.1~$10^{-5}$ & 3.2~$10^{-5}$ & 0.12 & 0.14 \\
10& 1.3~$10^{-4}$ & 1.4~$10^{-4}$ & 5.7~$10^{-2}$ &6.8~$10^{-2}$ \\
&&&&\\
\hline
\end{tabular*}

\end{document}